\documentclass[review]{elsarticle}

\usepackage{lineno,hyperref}
\modulolinenumbers[5]

\usepackage[english]{babel}
\usepackage{amsmath,amsthm,amsfonts,amssymb}
\usepackage[mathcal]{eucal}
\usepackage{lmodern}
\usepackage{geometry}
\usepackage{graphicx}
\usepackage{latexsym}
\usepackage{epsfig}
\usepackage{scalefnt}
\usepackage{dsfont}
\usepackage{hyperref}
\usepackage{braket}
\usepackage{adjustbox}
\usepackage{subfigure}
\usepackage{color}\newgeometry{left=2cm,right=2cm,top=2.5cm,bottom=2.5cm}

\journal{Physics Letters B}

\begin{document}

\begin{frontmatter}

\title{New class of solutions in the non-minimal O(3)-sigma model}

\author{F. C. E. Lima$^{a}$, and C. A. S. Almeida$^{a,}$\footnote{Correspondence to: Theoretical Physics Group, Department of Physics, Universidade Federal do Ceará, Campus do Pici, Fortaleza-CE, C. P. 6030, Brazil.\\ Email address: cleiton.estevao@fisica.ufc.br (F. C. E. Lima), carlos@fisica.ufc.br (C. A. S. Almeida).}}

\address[mymainaddress]{Universidade Federal do Cear\'{a} (UFC), Departamento de F\'{i}sica, Campus do Pici, Fortaleza-CE, C. P. 6030, 60455-760, Brazil.}

\begin{abstract}
For the study of topological vortices with non-minimal coupling, we built a kind of non-canonical O(3)-sigma model, with a Maxwell term modified by a dielectric function. Through the BPS formalism an investigation is made on possible configurations of vortices in topological sectors of the sigma model and the real scalar field. For a particular ansatz, the solutions of the topological sector of the real scalar field are described by the known kink solutions. On the other hand, when studying the vortices in non-minimal sector of the pure O(3)-sigma model, it is detected the emergence of solutions that generate solitary waves similar to structures derived from a KdV-like theory. We observed that in the study of mixed models, namely, the topological sector of the O(3)-sigma model coupled to the topological sector of the real scalar field, the vortex solutions assume a profile of a step function. Then, when kinks of the topological sector of the scalar field are interacting with the field of the sigma model, it makes the field solutions of the O(3)-sigma model become extremely localized, making the vortice structures non-physical.
\end{abstract}

%\begin{keyword}
%111111, 111111, 111111.
%\end{keyword}
\end{frontmatter}

%\linenumbers

\section{Introduction}

Symmetry is one of the most interesting properties of physics. This is due to the fact that symmetries are directly related to the conservation laws \cite{Hanc,Halder}. However, when we have models that admit a spontaneous symmetry breaking, a very interesting physical phenomenon occurs. This phenomenon is known as phase transitions \cite{Thouless,Anderson,Kosterlitz}. As a consequence of these transitions we have the emergence of topological structures in theory, for example, the vortices \cite{Nielsen,Schroers,Ghosh,Lee,Kim,LPA2}. Vortices are structures that emerge from theories in (2+1)D and are interesting for their similarity to the Abrikosov vortices studied in condensed matter physics \cite{Abrik}. 

The (2+1)D theories are interesting because they allow the existence of solitonic solutions in the theory \cite{LDA}. In general, solitons are objects of intense and growing interest from many researchers, both for fundamental reasons and for applicability \cite{Laine,Manton,Shnir,Ackerman,Kosevich}. In (1+1) dimensional spacetime, as seen in Ref. \cite{LDA}, the initial theory of solitons rests on the theory of equations that are fully integrable when using techniques such as inverse dispersion transformation. It is worth mentioning that the wave solutions of the solitons (in 3D)  that we will discuss throughout this work are not fully integrable even with the implementation of the well-known BPS technique that we will discuss next. 

It is evident that the O(3)-sigma model in 3D spacetime is well known and very popular in the theoretical physics scenario due to its notoriety in the study of topological structures or topological solitons \cite{Schroers,Lesse,PGhosh,PGhosh1,Casana1,Casana,CDC,FCCA}. In particular, this popularity of the O(3)-sigma model is due to the fact that when submitted to potentials that spontaneously break symmetry, it can become into a model with BPS properties facilitating its analysis. A negative point of the pure O(3)-sigma model arises when we look at the interpretation of the model from the point of view of a particle physicist \cite{Ghosh}. In short, the disadvantage is the scale invariance that these models can have, making them unsuitable as models for particle description. On the other hand, the O(3)-sigma model is intensively studied due to its applications, such as in the study of Heisenberg ferromagnetism \cite{Brezin,Haldane}.

In a quick review of the literature, we can start in 1995, when Schroers \cite{Schroers} examined the Bogomol'nyi solitons in a gauged O(3)-sigma model. In fact, Schroers shows that the scale invariance of the sigma model can be broken by measuring a subgroup U(1) of the O(3) symmetry. In other words, the scale invariance of the O(3)-sigma model can be broken by introducing a gauge field. Following Schoers' work, in 1996, Ghosh et. al.\cite{Ghosh} proposed the study of the O(3)-sigma model with a $\phi^4$ type potential and coupled to the Chern-Simons field. Next, several studies have been carried out investigating the topological structures in the O(3)-sigma model, as shown in Refs. \cite{muk,CCA,rothe,Casana1}.

In general, to investigate classical field solutions in 3D spacetime it is necessary to use some approach. In summary, it is common to use the Bogomol'nyi-Prasad-Sommefield (BPS) technique to study the configurations of self-dual field of models that have BPS property \cite{B,PS}. In fact, to know the classical solutions of a domain wall \cite{TLee,Zeldovich}, or vortex lines \cite{Nielsen}, the stability is achieved by topologically selecting non-trivial boundary conditions that lead to the existence of conserved quantities for solutions, quantities that play a role analogous to charge \cite{B,Yu,Monas}. Thus, in topological models with BPS property, the conserved amount is called topological charge and in the BPS limit this amount is equivalent to the energy density of the structure.

Recently, new class of solutions and structures has been studied \cite{Casana3}. In particular, vortices with new characteristics are found stimulating the extension of the U(1) symmetry \cite{Witten,Peterson}. Other interesting examples can arise when we consider models with dielectric permeability deriving from a quantum correction theory \cite{LPA}. In this case, we lose the freedom to define the form of the dielectric function of the model. Other generalized theories have been intensively studied seeking a better understanding of the dynamics and physical properties of structures \cite{LN,Babi,LA}.

Several studies were performed on the topological structures of O(3)-sigma model gauged by Maxwell \cite{FCCA} and Chern-Simons fields \cite{LDA,CDC}. In general, these studies demonstrate the existence of vortex structures in O(3)-sigma models. However, the question arises: what influence can a scalar field have on a topological sector (i. e., on the structures) of the O(3)-sigma model? Seeking to answer this question, we motivate our study to understand the influence of a topological sector of an additional scalar field at the structures of the O(3)-sigma model.

In this work, we found new class of solutions that describe the topological vortices of a generalized model with non-minimal coupling and governed by the fields of the O(3)-sigma model with non-canonical dynamics. The model is characterized also by a real scalar field, and by Maxwell field coupled to a dielectric function. We show that this new class of solutions depends on the form of the kinetic term of the O(3)-sigma model.

Our work is organized as follows: In Sec. II, we propose a generalized Lagrangian density that generates a possible scheme for the emergence of new class of structures. In Sec. III, using the BPS technique and postulating that the vortices are spherically symmetrical, it is performed the study of possible cases, namely, the canonical case, the generalized pure case and the mixed case one. In Sec. IV, we present the conclusions and final considerations.

%------------------------------------------------------------------------
%\section{Possible scheme for the emergence of new class of structures}
\section{Generalized non-minimal O(3)-sigma model}
 
Let us begin our investigation in a flat spacetime of $(2+1)$-dimensions. We are motivated by Ref. \cite{Casana} and we are aware that generalized models can give origin to structures with a considerable flux of radiated energy. Finally, we also seek the possibility of arising of internal structure. We propose the study of the O(3)-sigma model with non-minimal coupling and coupled to a scalar field described by the following Lagrangian density
 \begin{align}\label{model}
     \mathcal{L}=\frac{\mathcal{F}(\Phi, \psi)}{2}\nabla_{\mu}\Phi\cdot \nabla^\mu\Phi-\frac{\mathcal{G}(\Phi,\psi)}{4}F_{\mu\nu}F^{\mu\nu}+\frac{1}{2}\partial_\mu \psi\partial^\mu \psi-\mathcal{V}(\phi_3,\psi),
 \end{align}
where $\mathcal{V}(\phi_3,\psi)$ is a potential that breaks the symmetry of the model. Here, it is important to mention that the $\Phi$ field is a triplet of scalar fields that respects the constraint of the O(3)-sigma model. The $\psi$ field is a real scalar field. Generalized models were considered for the first time by Lee and Nam, in a Chern-Simons-Higgs model \cite{LN}. However, after the paper of Lee and Nam, many works have studied generalized models \cite{LPA,LA,FCCA1}. In our work, the function $\mathcal{F}(\Phi,\psi)$ is an arbitrary function used to modify the contribution of the kinetic term \cite{FCCA1}. On the other hand, the function $\mathcal{G}(\Phi,\psi)$ is an arbitrary function, known as the dielectric permeability function. In fact, the function $\mathcal{G}(\Phi, \psi)$ is responsible for adjusting the contribution of the electromagnetic field \cite{LPA,LN}. In this model, it is natural to define these generalization functions as being positive-defined due to the symmetry of the O(3)-sigma model.

The usual covariant derivative is defined as
\begin{align}
    D_\mu\Phi=\partial_\mu\Phi+eA_\mu \hat{n}_3\times\Phi.
\end{align}
However, we are interested in the study of the non-minimal model \cite{CCA}. In this way, the covariant derivative is now defined by
 \begin{align}
     \nabla_\mu \Phi=\partial_\mu\Phi+\bigg(eA_\mu+\frac{g}{2}\varepsilon_{\mu\nu\lambda}F^{\nu\lambda}\bigg)\hat{n}_3\times\Phi.
 \end{align}
 
Throughout this work, some conventions are adopted, namely, the metric signature $\eta_{\mu\nu}=$diag$(+,-,-)$, and the electromagnetic tensor is $F_{\mu\nu}=\partial_{\mu}A_ {\nu}-\partial_{\nu}A_{\mu}$.
 
 The equation of motion for the gauge field is 
 \begin{align}\label{EqGauge}
     \partial_\nu [g\varepsilon_{\mu\nu\lambda}\mathcal{F}(\Phi,\psi)\hat{n}_3\cdot(\Phi\times\nabla^\mu\Phi)-\mathcal{G}(\Phi,\psi)F^{\nu\lambda}]=j^\nu.
 \end{align}
Observing the Eq. (\ref{EqGauge}), it is visible in the absence of the generalization, that the gauge field equation is similar to the equation presented in Refs. \cite{CCA,Almeida1} (in the absence of the Chern-Simons field).

The matter field current is defined as
\begin{align}
    j^\nu=e \,\hat{n}_3\cdot (\Phi\times\nabla^\nu\Phi),
\end{align}
where $\textbf{J}^\nu=-j^\nu\cdot\hat{n}_3$.

Analyzing Gauss's law, i. e., the $\nu=0$ component of Eq. (\ref{EqGauge}), it is convenient to consider $A_0=0$. In this case, the structures are  purely magnetic.

Similarly, the equation of motion of the matter field $\Phi$ is described by
\begin{align}
    \mathcal{F}(\Phi,\psi)\nabla_\mu\nabla^\mu\Phi=\frac{\mathcal{F}_{\Phi}}{2}\nabla_\mu\Phi\cdot\nabla^\mu\Phi-\frac{\mathcal{G}_\Phi}{4}F_{\mu\nu}F^{\mu\nu}-\mathcal{V}_\Phi,
\end{align}
where $\mathcal{F}_{\Phi}=\frac{\partial\mathcal{F}}{\partial\Phi}$, $\mathcal{G}_{\Phi}=\frac{\partial\mathcal{G}}{\partial\Phi}$, and $\mathcal{V}_{\Phi}=\frac{\partial\mathcal{V}}{\partial\Phi}$.

On the other hand, the equation of the real scalar field is given by
\begin{align}
    \partial_\mu\partial^\mu\psi=-\frac{\partial\mathcal{V}}{\partial\psi}.
\end{align}

To study the finite energy configurations, we turn our attention to the construction of the energy functional of the model. Using the $T_{00}$ component of the energy-momentum tensor, we arrive in
\begin{align}\label{funcional1}
    E=\frac{1}{2}\int\, d^2x \bigg[\mathcal{F}\nabla_{i}\Phi\cdot\nabla^i\Phi+(\partial_i\psi)^2+\mathcal{G}F_{ij}F^{ij}+2\mathcal{V}\bigg].
\end{align}

To inspection the BPS property of the model, it is convenient to rewrite the potential in terms of an arbitrary function. This function is called superpotential \cite{Vachaspati}. In summary, the potential is obtained in terms of a superpotential when the energy is saturated. These superpotentials are used in several scenarios to study the first-order formalism of theory (see Refs. \cite{Moreira,ca,Zh}). To implement the first-order formalism and obtain the self-dual equations, let us consider two superpotentials of the type $U(\Phi)=U(\phi_3)$ and $W=W(\psi)$. These superpotentials will relate to $\mathcal{V}$ at the energy saturation limit. We remark that the only constraint on these superpotentials is that they must produce the spontaneous breaking of symmetry in theory, as we will see in Eq. (\ref{v}).
%These superpotentials will relate to $\mathcal{V}$ at the energy saturation limit, as we will see in Eq. (\ref{v}). 

Without losing the generality, we restructured the energy as follows:
\begin{align}\nonumber
    E=& \int \, d^2x\, \bigg\{\bigg[\frac{\mathcal{F}}{2}\bigg(\nabla_i\Phi\mp\frac{1}{\mathcal{F}^{1/2}}\varepsilon_{ij}\Phi\times\nabla_j\Phi\bigg)^2+\frac{\mathcal{G}}{2}\bigg(F_{ij}\pm\sqrt{\frac{2U}{\mathcal{G}}}\bigg)^2+\frac{1}{2}\bigg(\partial_i\psi\mp \frac{W_{\psi}}{r}\bigg)^2\bigg]+\\
    &\pm\mathcal{F}^{1/2}\varepsilon_{ij}\Phi\cdot (\nabla_i \Phi\times\nabla_j \Phi)\mp F_{ij}\sqrt{2U\mathcal{G}}\pm\frac{W_\psi \psi'}{r}-\frac{W_{\psi}^{2}}{2r}-U+\mathcal{V}\bigg]\bigg\},
\end{align}
where $W_\psi=\frac{\partial W}{\partial\psi}$.

For the model to admit BPS property, let us assume that 
\begin{align}\label{v}
    \mathcal{V}(\phi_3,\psi)=U+\frac{W_{\psi}^{2}}{2r}.
\end{align}

By inspection, the conserved topological current of sector $\Phi$, i. e., of the O(3)-sigma model is
\begin{align}
    \mathcal{J}_\mu^{(\sigma)}=\frac{1}{8\pi}\varepsilon_{\mu\nu\lambda}\bigg[\mathcal{F}^{1/2}\Phi\cdot\nabla^\nu\Phi\times\nabla^\lambda\Phi-F^{\nu\lambda}\sqrt{2U\mathcal{G}}\bigg].
\end{align}

On the other hand, the current-like tensor in the sector of $\psi$ scalar field is
\begin{align}
    \mathcal{J}^{(\psi)}_{\mu\nu}=\frac{1}{\sqrt{2\pi}}\varepsilon_{\mu\nu\alpha}\partial^{\alpha}\psi,
\end{align}
so the total topological current of the model is described by 
\begin{align}
    \mathcal{J}_{top}=\mathcal{J}_{top}^{(\sigma)}+\mathcal{J}_{top}^{(\psi)}.
\end{align}

The expression of energy can be simplified by rewriting it as follows: 
\begin{align}\nonumber
    E=& \int \, d^2x\, \bigg\{\bigg[\frac{\mathcal{F}}{2}\bigg(\nabla_i\Phi\mp\frac{1}{\mathcal{F}^{1/2}}\varepsilon_{ij}\Phi\times\nabla_j\Phi\bigg)^2+\frac{\mathcal{G}}{2}\bigg(F_{ij}\pm\sqrt{\frac{2U}{\mathcal{G}}}\bigg)^2+\frac{1}{2}\bigg(\partial_i\psi\mp \frac{W_\psi}{r}\bigg)^2+\\
    \pm&\int\, d^2x\, \mathcal{E}_{BPS}, \label{en1}
\end{align}
where, the BPS energy density is written in terms of the topological charge, namely,
\begin{align}
\mathcal{E}_{BPS}=\mathcal{Q}^{(\sigma)}+\mathcal{Q}^{(\psi)}.
%\pm\bigg[\mathcal{F}^{1/2}\varepsilon_{ij}\Phi\cdot (\nabla_i \Phi\times\nabla_j \Phi)+ F_{ij}\sqrt{2U\mathcal{G}}-\frac{W_\psi \psi'}{r}\bigg].
\end{align}
In terms of field variables, generalization functions, and vorticity of the structure, the BPS energy density is shown in Sec. 2.1 (see Eq. (\ref{ec}) ahead).

From Eq. (\ref{en1}), it is evident that the energy of static field configurations is limited from below. Therefore, at the energy saturation limit, we obtain the first order differential equations, namely,
\begin{align}\label{bps}
    &\mathcal{F}^{1/2}\nabla_i\Phi=\pm\varepsilon_{ij}\Phi\times\nabla_j\Phi;\\ \label{bps2}
    &F_{12}=\mp\sqrt{\frac{2U}{\mathcal{G}}};\\ \label{bps3}
    &\partial_i\psi=\pm\frac{W_\psi}{r}, 
\end{align}

\subsection{The spherically symmetrical vortices}

As proposed in Refs. \cite{Schroers,Ghosh}, in order to investigate the solutions that are invariant under simultaneous rotations and reflection in spacetime and target space, we assume that the behavior of the O(3)-sigma respects the spherically symmetric ansatz, i. e., 
\begin{align}\label{ans1}
    \Phi(r,\theta)=\begin{pmatrix}
    \sin{f(r)}\cos{N\theta}\\
    \sin{f(r)}\sin{N\theta}\\
    \cos{f(r)}
    \end{pmatrix}.
\end{align}
For the real scalar field $\psi$, we assume that 
\begin{align}\label{ans2}
    \psi=\psi(r).
\end{align}

Once again, following the method proposed by Schroers in Ref. \cite{Schroers} and Ghosh \cite{Ghosh}, let us assume that the gauge field is
\begin{align}\label{ans3}
    \textbf{A}=-\frac{Na(r)}{er}\hat{e}_{\theta},
\end{align}
where $N$ is the winding number of the vortex. In fact, these behaviors (\ref{ans1}) and (\ref{ans3}) are useful for the study of purely magnetic structures. Interested in this type of structure, we turn our attention to the study of equations (\ref{bps}-\ref{bps3}), which in terms of variable fields are
\begin{align} \label{bps4}
    &f'(r)=\pm\frac{N}{r\sqrt{\mathcal{F}(f(r),\psi)}}[a(r)-1]\sin{f(r)}, \\ \label{bps5}
    &a'(r)=\pm\frac{er}{N}\sqrt{\frac{2U}{\mathcal{G}(f(r),\psi)}}, \\ \label{bps6}
    &\psi'(r)=\pm\frac{1}{r}\frac{\partial W}{\partial\psi}.
\end{align}

Note that, to write Eq. (\ref{bps5}), we observe that $F_{12}=-\vert\vert\textbf{B}\vert\vert$, and that the magnetic field of the vortex is given by
\begin{align}\label{cmagnet}
    \textbf{B}=\nabla\times\textbf{A}\equiv \frac{Na'(r)}{er}\hat{e}_{\phi}\to F_{12}=-B=-\vert\vert\textbf{B}\vert\vert=-\frac{Na'(r)}{er},
\end{align}
where $a'(r)$ is positive for all $r\geq 0$. The condition previously mentioned is because the gauge field couples to the O(3)-sigma field. In the O(3)-sigma sector, the field variable is defined for $r\geq 0$.

These structures produce a quantized magnetic flux $\Phi_B$ given by 
\begin{align}\label{f111}
    \Phi_B=\int_{surface}{\bf B}\cdot d{\bf S}=-\frac{2\pi N}{e}[a(\infty)-a(0)].
\end{align}
To obtain the magnetic flux, we consider the planar characteristic of the vortices so that the surface for the calculation of the integral (\ref{f111}) must have cylindrical symmetry.% The upper basis of this cylindrical surface is above the magnetic vortex.}

For the ansatz (\ref{ans1}) and (\ref{ans2}), the BPS energy in terms of the field variables is
\begin{align}\label{ec}\nonumber
     E_{BPS}=&\int\, d^2x\, \mathcal{E}_{BPS}=\int\, d^2x\, (\mathcal{Q}^{(\sigma)}+\mathcal{Q}^{(\psi)})\\
     =&\pm\int\, \bigg[\mathcal{F}^{1/2}\frac{N(1-a(r))}{r}f'(r )\sin{f(r)}+\frac{Na'(r)}{er}\sqrt{2U\mathcal{G}}+\frac{W_{\psi}\Psi'(r)}{r}\bigg]\, d^2x,
\end{align}
so that the topological charge is the integrand of the Eq. (\ref{ec}).

At this point, it is convenient to consider a set of boundary conditions for the fields $\Phi$, $\textbf{A}$ and $\psi(r)$, so that we can investigate the topological structures of the model. In this way, let us assume the following conditions: 
\begin{align}\label{cont}
    &f(r\to 0)\to 0, \, \, \, \, \, f(r\to \infty)\to \pi, \\ \label{cont1}
    &a(r\to 0)\to 0, \, \, \, \, \, a(r\to \infty)\to -\eta_1,\\ \label{cont2}
    &\psi(r\to 0)\to -1, \, \, \, \, \, \psi(r\to \infty)\to 1.
\end{align}

This set of conditions leads us to structures with a magnetic flux given by
\begin{align}
    \Phi_B=\frac{2\pi N\eta_1}{e}.
\end{align}
Therefore, as we had previously assumed, the vortices studied throughout this work will be purely magnetic, since when analyzing Gauss's law we assume that $A_0=0$. 

\section{Possible scheme to generate new structures in the O(3)-sigma model}

To begin this topic, allow us to motivate our discussion by Ref. \cite{Liao}. Their authors show that a model governed by two scalar fields in a $\phi^4$-like theory generates deformable kink-like structures. Also, they observed that when the interaction of the two topological sectors is similar, the kink undergoes a contraction. This contraction may induce the appearance of multi-kink structures in one of the sectors. To build an extension of this theory, let us promote a scalar field for a vector field, i. e., the O(3)-sigma model, and the second one will still be a scalar field. In this case, the interactions are
\begin{align}\label{PotE}
    \mathcal{W}=\alpha\psi-\frac{1}{3}\alpha\psi^3 \, \, \, \, \, \text{and} \, \, \, \, \, U=\beta(\hat{n}_3\cdot\Phi)-\frac{1}{3}\beta(\hat{n}_3\cdot\Phi)^3,
\end{align}
with $\alpha$ and $\beta$ are constants. For the special potentials assumed in (\ref{PotE}), the effective potential is given by
\begin{align}
    \mathcal{V}=\frac{\alpha}{2r}(1-\psi^2)+\beta\cos{f(r)}\bigg(1-\frac{1}{3}\cos^2 f(r)\bigg).
\end{align}

In this way, the BPS equations are
\begin{align}\label{bps7}
    &f'(r)=\pm\frac{N}{r\sqrt{\mathcal{F}}}[a(r)-1]\sin{f(r)},\\ \label{bps8}
    &a'(r)=\pm\frac{\sqrt{2}er\beta}{N}\bigg[\cos f(r)-\frac{1}{3}\cos^3 f(r)\bigg],\\ \label{bps9}
    &\psi'(r)=\pm\frac{\alpha}{r}(1-\psi^2).
\end{align}
To simplify the calculations, we restrict our study to the case $\mathcal{G}=U^{-1}$. In this case, the dielectric permeability of the environment behaves according to the profile of $U^{-1}$, as discussed in Refs. \cite{LPA,DMM}. This kind of choice leads us to certain class of topological solitons \cite{Adam}. In fact, choosing the condition $\mathcal{G}=U^{-1}$ frees us from dependence on $\mathcal{G}$ in the BPS equations (see our BPS equations). Still in this context, see the discussion by Adam and collaborators in Ref. \cite{Adam}, where for the study of incompressible solitons, the generalization function of the kinetic term is the inverse of the potential.

To obtain the BPS equations independent of the constants $\beta$ and $e$, one can assume without loss of generality that $\beta=1/\sqrt{2}e$. In fact, this is equivalent to saying that the proportionality constant of the O(3)-sigma model interaction is inversely proportional to $\sqrt{2}e$. This consideration leads us to independent equations for $e$ and $\beta$, namely,
\begin{align}\label{bps10}
    &f'(r)=\pm\frac{1}{r\sqrt{\mathcal{F}}}[a(r)-1]\sin{f(r)},\\ \label{bps11}
    &a'(r)=\pm r\bigg[\cos f(r)-\frac{1}{3}\cos^3 f(r)\bigg],\\ \label{bps12}
    &\psi'(r)=\pm\frac{\alpha}{r}(1-\psi^2).
\end{align}
The choice of the constant $\beta$ is to avoid loading additional terms that will not influence the profile of the topological structure. Similar considerations are used in Ref. \cite{CCA}. In agreement with the existing works \cite{Schroers,Ghosh,Lee,Kim}, we assume the winding number $N=1$.

Manipulating Eqs. (\ref{bps10}) and (\ref{bps11}), we can write the following equation for the variable field $f(r)$:
\begin{align}
    f''(r)-\frac{f'(r)^2}{\tan f(r)}+f'(r)\bigg(\frac{1}{r}+\frac{1}{2\mathcal{F}}\frac{d\mathcal{F}}{dr}\bigg)+\frac{\sin 2f(r)}{2\sqrt{\mathcal{F}}}\bigg(\frac{\cos^2 f(r)}{3}-1\bigg)=0.
\end{align}

\begin{figure}[ht!]
    \centering
    \includegraphics[height=5cm,width=7.5cm]{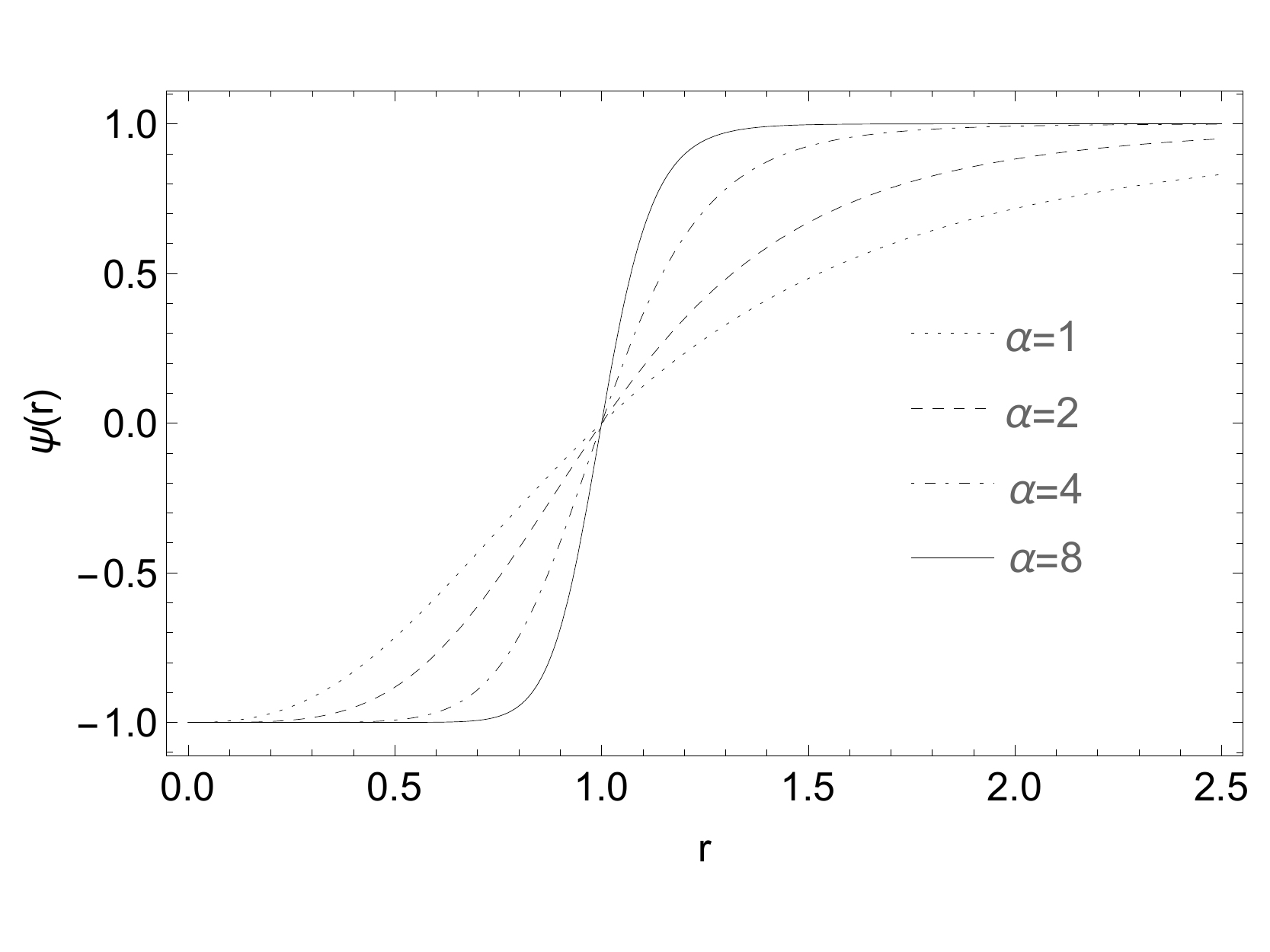}%\hspace{1cm}
    \includegraphics[height=5cm,width=7.5cm]{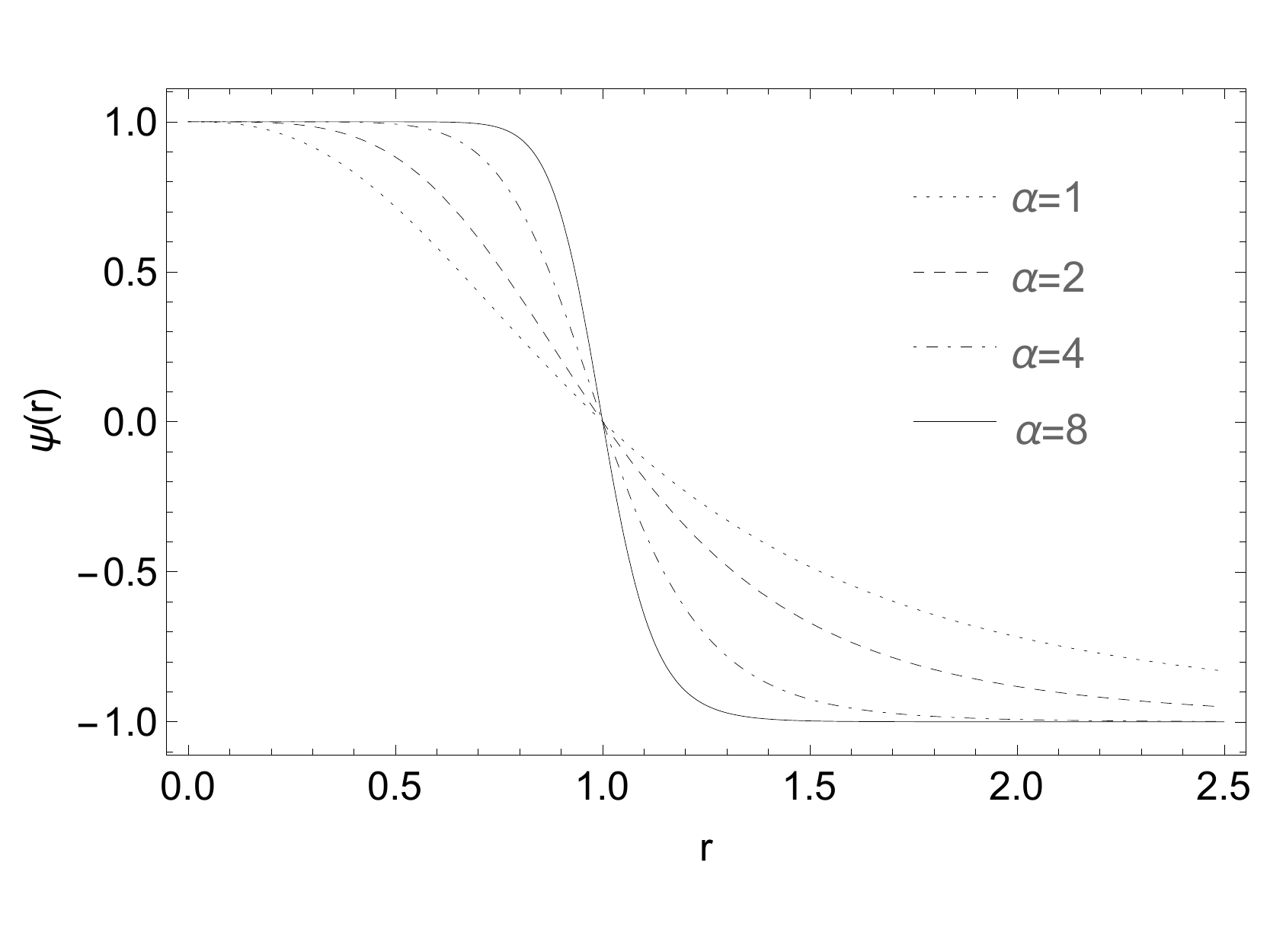}\\ \vspace{-0.5cm}
    \hspace{1cm}(a) \hspace{7.5cm} (b)
    \vspace{-0.5cm}
    \caption{(a) Kink solutions of the field $\psi(r)$ for several values of $\alpha$. (b) Antikink solutions of the field $\psi(r)$ for several values of $\alpha$.}
    \label{fig1}
\end{figure}

On the other hand, analyzing the Eq. (\ref{bps12}), it is observed that the solution for $\psi(r)$ is
\begin{align}
    \psi(r)=\pm\frac{r^{2\alpha}-r_{0}^{2\alpha}}{r^{2\alpha}+r_{0}^{2\alpha}},
\end{align}
where $r_0$ it is a constant of integration. For convenience, let us assume $r_0=1$, which allows us to write the field $\psi(r)$ as
\begin{align}
    \psi(r)=\pm\tanh(\text{ln}(r^\alpha)).
\end{align}
Therefore, the topological solutions of the field $\psi(r)$ are the so-called kink (positive sign) and antikink (negative sign) solutions. The parameter $\alpha$ will be responsible for contracting these structures. The topological solutions of $\psi(r)$ are shown in Fig. \ref{fig1}. With the illustration in Fig. \ref{fig1} it is clear that topological solutions that satisfy the boundary condition (\ref{cont2}) are the kink solutions described by
\begin{align}\label{psi}
    \psi(r)=\tanh(\text{ln}(r^\alpha)).
\end{align}

In order to investigate the emergence of new class of topological solutions in the O(3)-sigma model, we must define a profile for the function that generalizes the non-canonical term of the model, i. e., the function $\mathcal{F}(\phi_3,\psi)$. 

\subsection{The canonical model: $\mathcal{F}=1$}

In this case, the equations (\ref{bps10}) and (\ref{bps11}) are reduced to
\begin{align} \label{bps14}
    &f'(r)=\pm\frac{1}{r}[a(r)-1]\sin{f(r)} \, \, \, \, \, \text{and} \, \, \, \, \, 
    a'(r)=\mp r\cos{f(r)}\bigg[\frac{1}{3}\cos^2 f(r)-1\bigg].
\end{align}

\begin{figure}[ht!]
    \centering
    \includegraphics[height=5cm,width=7.5cm]{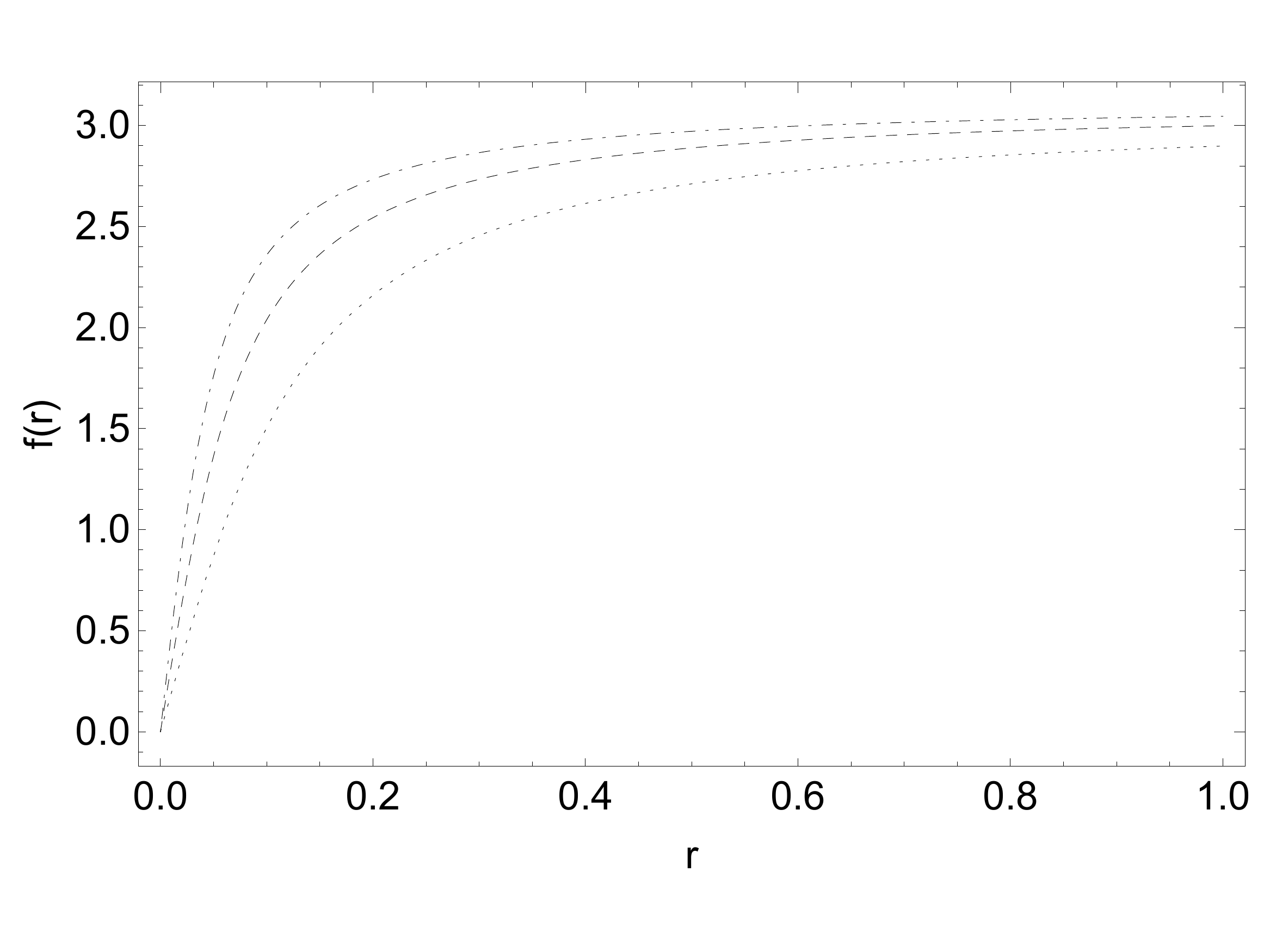}%\hspace{1cm}
    \includegraphics[height=5cm,width=7.5cm]{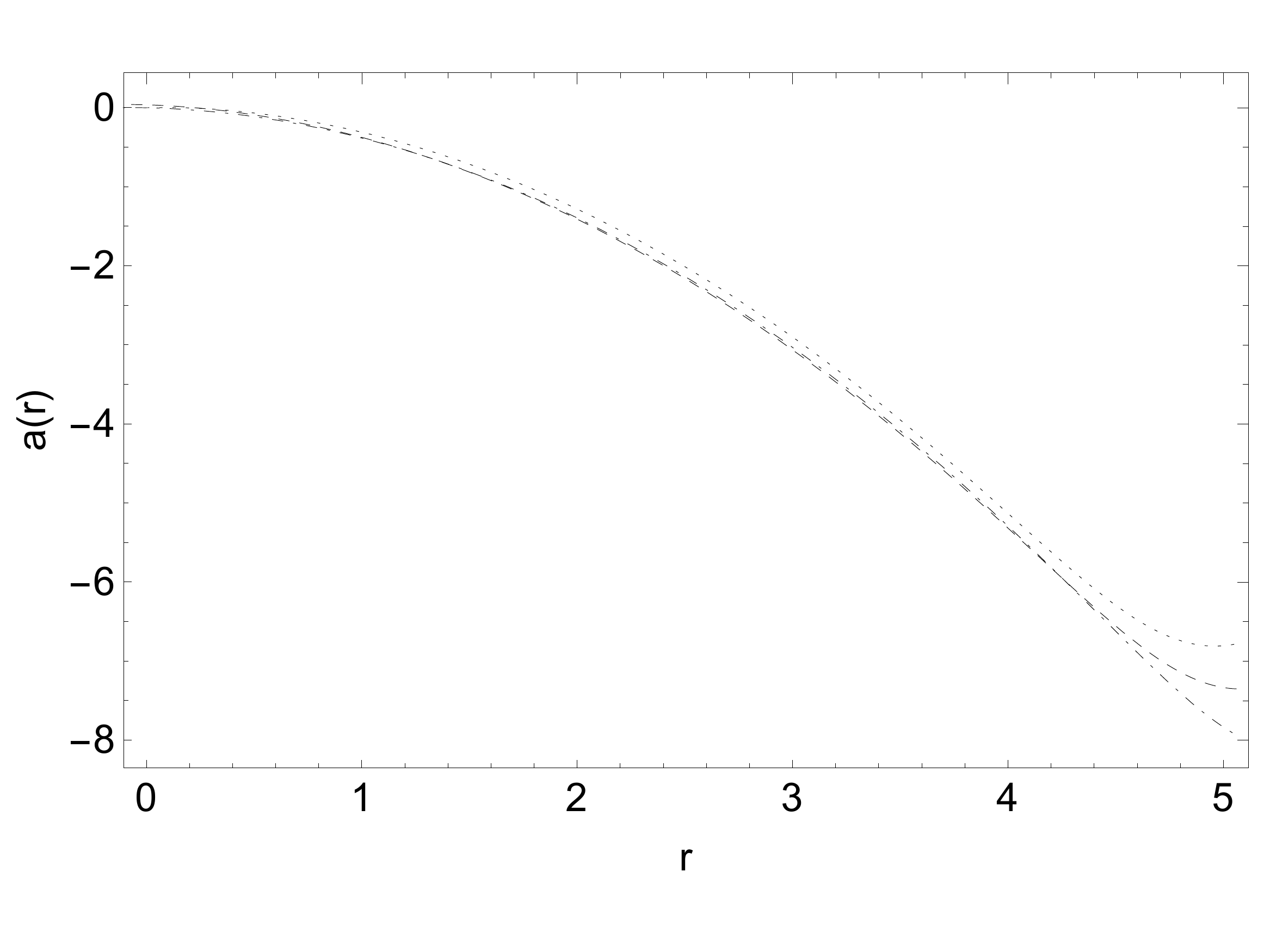}\\ \vspace{-0.5cm}
    \hspace{1cm}(a) \hspace{7.5cm} (b)
    \vspace{-0.5cm}
    \caption{(a) Solutions of the variable field $f(r)$ for vortices with winding number $N=1$. (b) Solution of the variable fields $a(r)$ associated to the gauge field when $N=1$.}
    \label{fig2}
\end{figure}

It is necessary to use a numerical technique to examine the solutions of Eq. (\ref{bps14}). Let us inspect the field configuration of the expression (\ref{bps14}) using the interpolation technique. Applying this approach, we find the solutions shown in Fig. \ref{fig2} for the vortices in the topological sector of the O(3)-sigma model.
 
\begin{figure}[ht!]
    \centering
    \includegraphics[height=5cm,width=7.5cm]{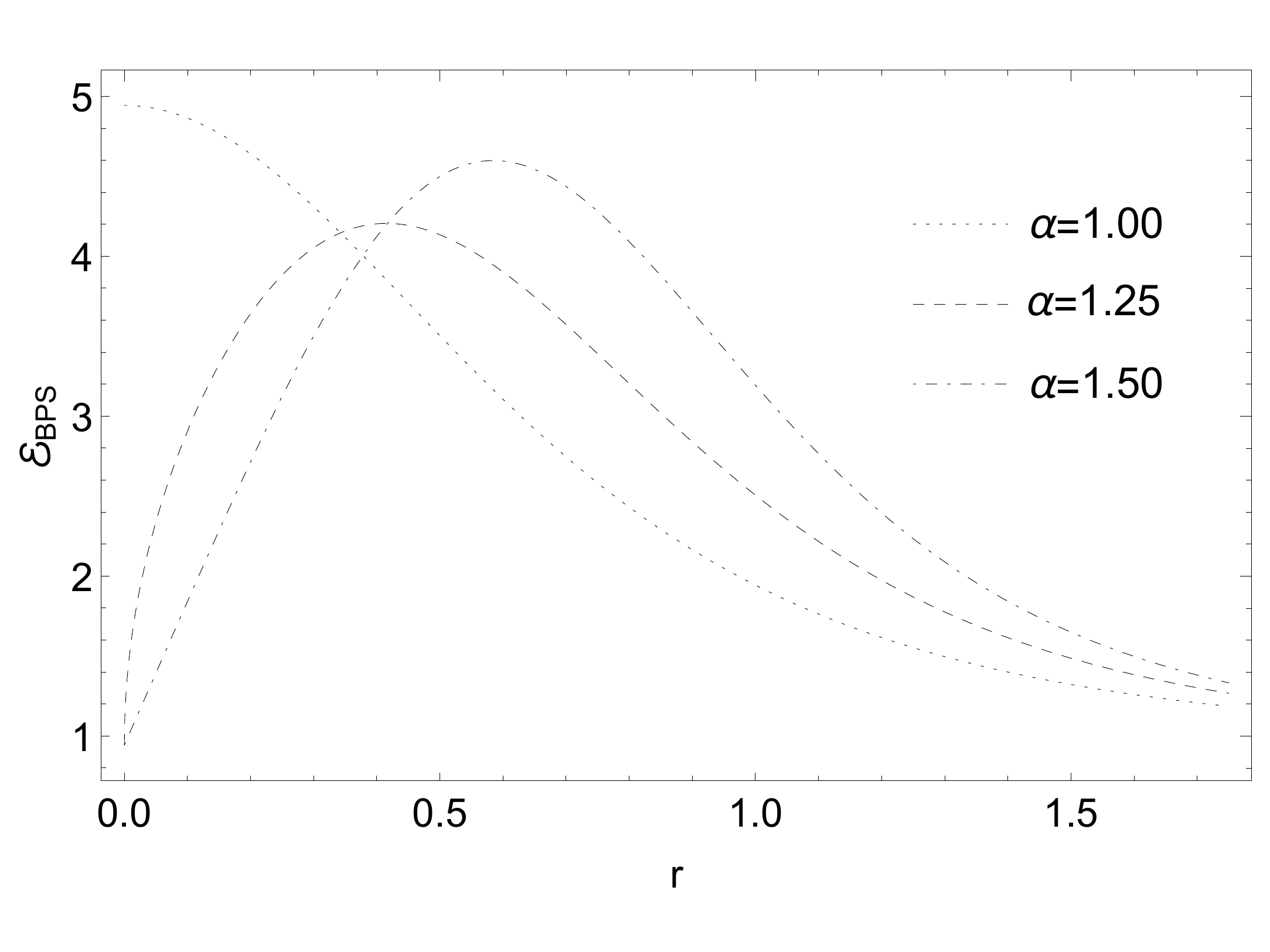}
    \includegraphics[height=5cm,width=7.5cm]{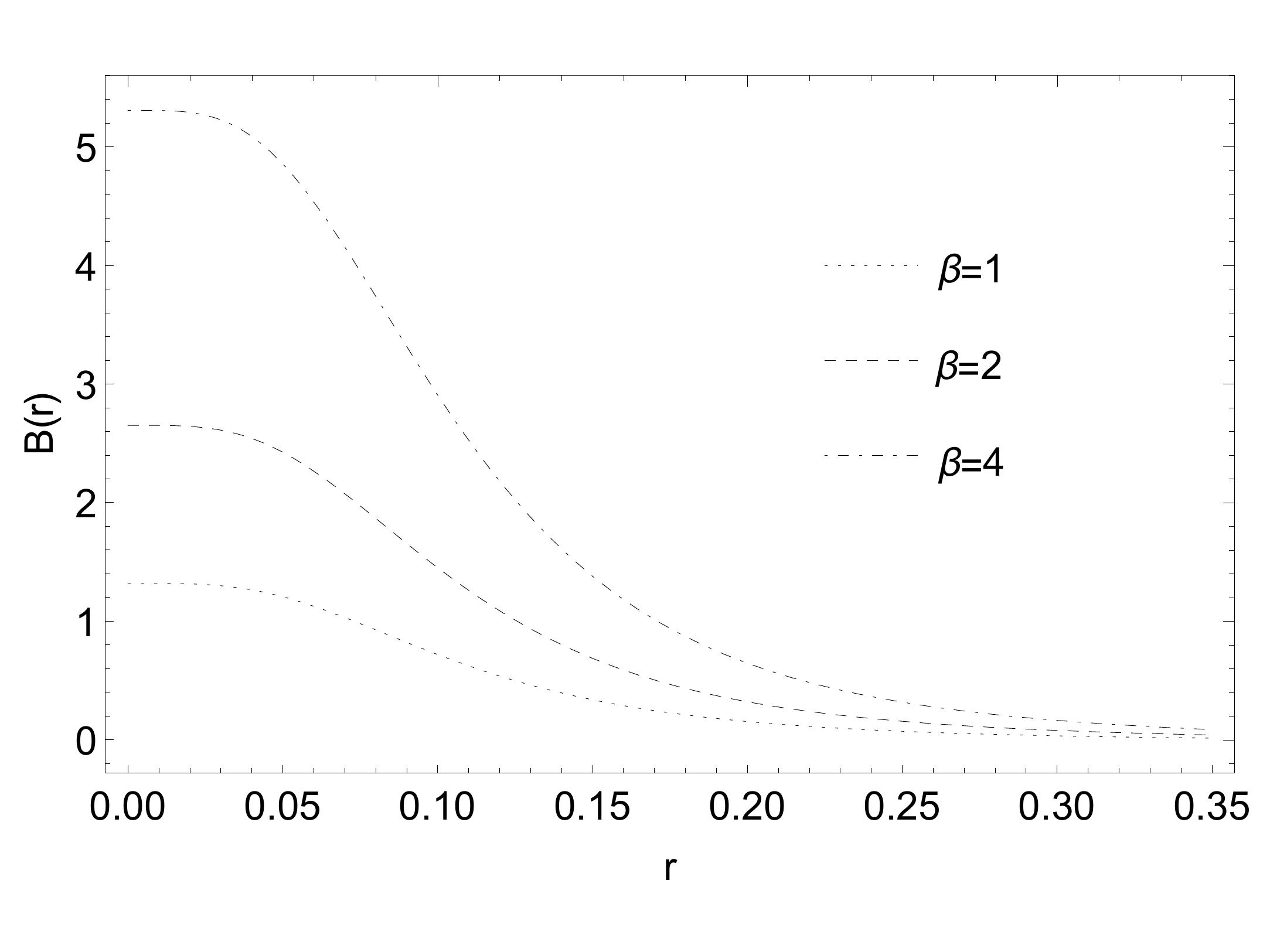}\\ \vspace{-0.5cm}
    \hspace{1cm}(a) \hspace{7.5cm} (b) \vspace{-0.5cm}
    \caption{(a) Energy density for several values of $\alpha$. (b) Vortex magnetic field for several values of $\beta$.}
    \label{fig3}
\end{figure}

An attractive behavior emerges from the investigation of the energy density and magnetic field of topological vortices without generalization (see Fig. \ref{fig3}). In this case, due to the particular choice of the potentials (\ref{PotE}), the solitons governed by canonical dynamics have an energy density related to the parameter $\alpha$. In fact, the parameter $\alpha$ is responsible for the symmetry breaking in the sector of the topological value of $\psi$. For higher values of $\alpha$, the energy density will tend to stay placed around the center of the kink described by $\psi(r)$. On the other hand, we note that the parameter $\beta$ is responsible for the symmetry breaking of the O(3)-sigma model. The parameter $\beta$ is also the only parameter which is responsible for changing the vortex's magnetic field, modifying the intensity of the magnetic flux. We remark that the parameters $\alpha$ and $\beta$ can be adjusted to obtain non-minimally couple structures similar to the configurations with minimal coupling.

\subsection{The pure non-canonical model: $\mathcal{F}=\phi_{3}^{m}$}

Due to the generalized model profile, we observe that there are three distinct possibilities to study our model with non-canonical dynamics. Among these, we have the pure case defined when the generalization of the kinetic term is only a function of $\phi_3$ or $\mathcal{F}=h(\phi_3)$. The mixed-case is defined by $\mathcal{F}=p(\psi)$ (in this case, we have a mixture of the kinetic term with the field $\psi$). The most general case would be when the generalization $\mathcal{F}$ is simultaneously a function of the fields $\phi_3$ and $\psi$. Let us now turn our attention to the study of the pure case. By convenience, allow us to consider the simplest possible case, namely,
\begin{align}\label{pure}
    \mathcal{F}(\phi_3)=\phi_{3}^{m}\equiv \cos^{m}{f(r)},
\end{align}
where $m$ is a positive integer.

In this case, the BPS equations in the topological sector of the O(3)-sigma model are
\begin{align}\label{bps15}
    &f'(r)=\pm\frac{1}{r}[a(r)-1]\sin{f(r)}\cos^{-\frac{m}{2}}f(r),\\ \label{bps16}
    &a'(r)=\pm r\bigg[\cos{f(r)}-\frac{1}{3}\cos^3f(r)\bigg].
\end{align}

\begin{figure}[ht!]
    \centering
    \includegraphics[height=5cm,width=7.5cm]{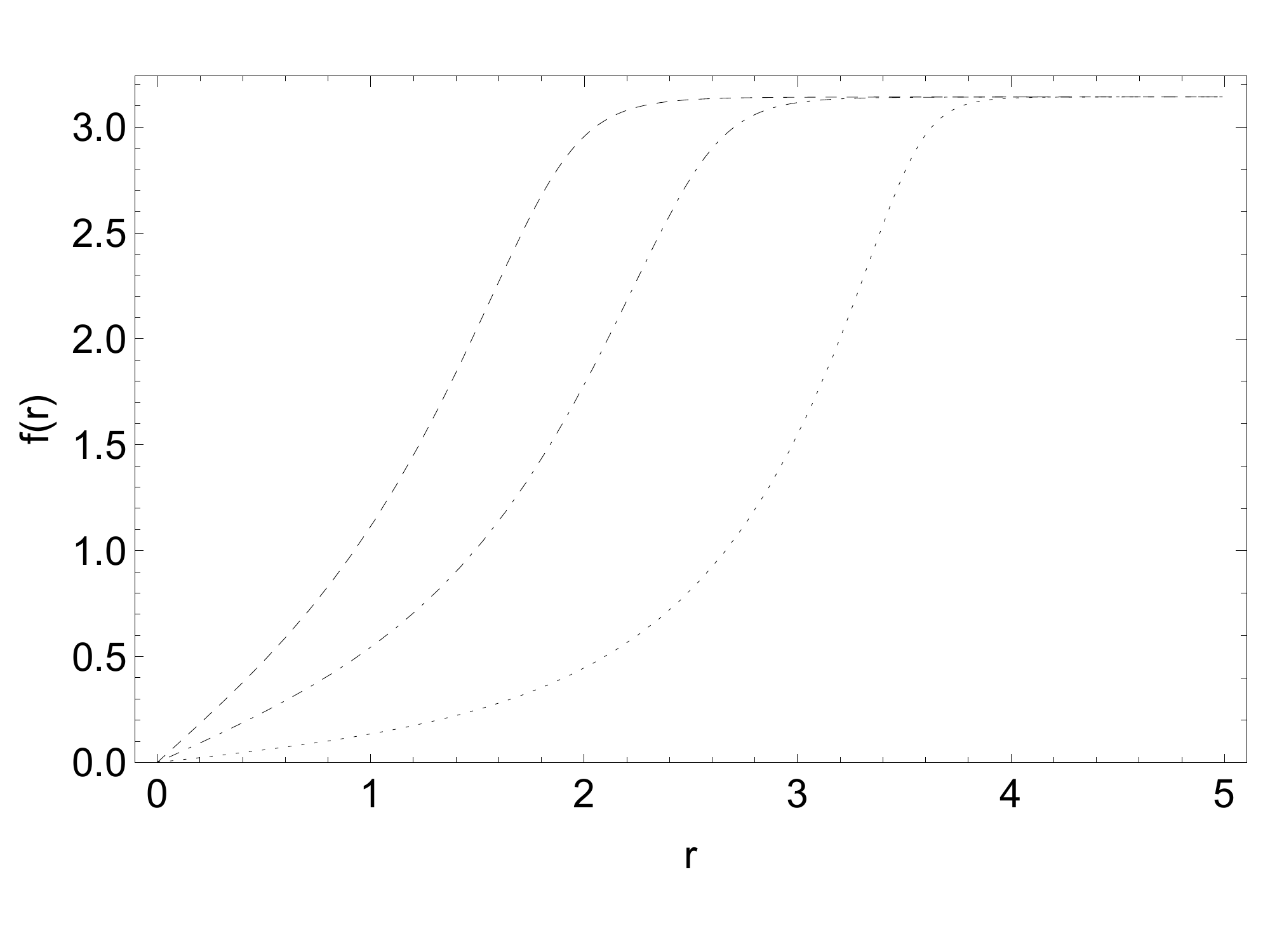}%\hspace{1cm}
    \includegraphics[height=5cm,width=7.5cm]{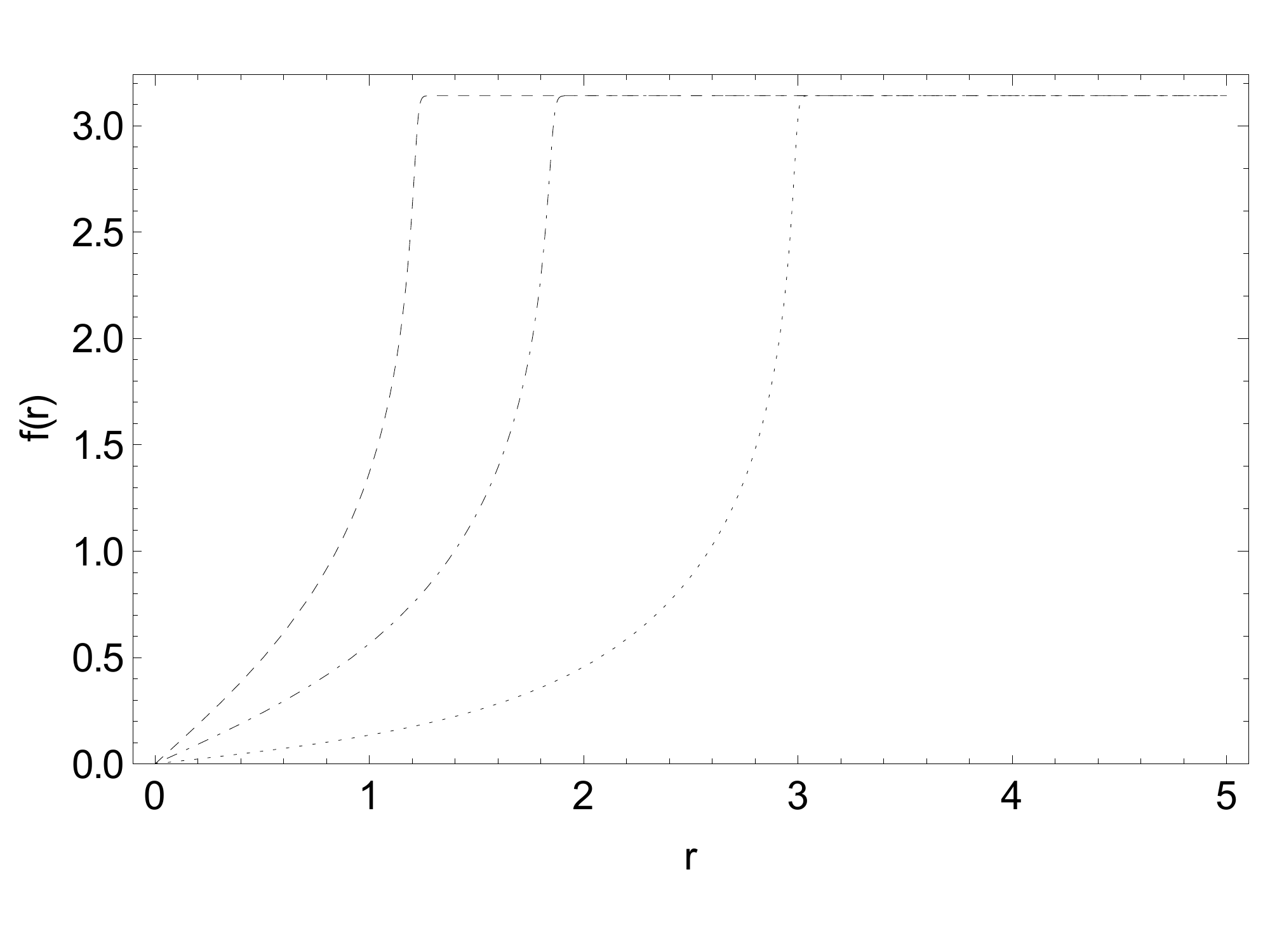}\\ \vspace{-0.5cm}
    \hspace{1cm}(a) \hspace{7.5cm} (b)
    \vspace{-0.5cm}
    \caption{(a) Numerical solution of the variable field $f(r)$ with winding number $N=1$ and $m=1$. (b) Numerical solution of the variable field $f(r)$ with winding number $N=1$ and $m=2$.}
    \label{fig4}
\end{figure}

As seen in the case without generalization (Sec. 3.1), the solutions of Eqs. (\ref{bps15}) and (\ref{bps16}) must be investigated numerically. Let us use a numerical interpolation approach and consider the topological boundary conditions (\ref{cont}-\ref{cont1}). The numerical solution for two distinct values of $m$ is shown in Fig. \ref{fig4}. 

\begin{figure}[ht!]
    \centering
     \includegraphics[height=5cm,width=7.5cm]{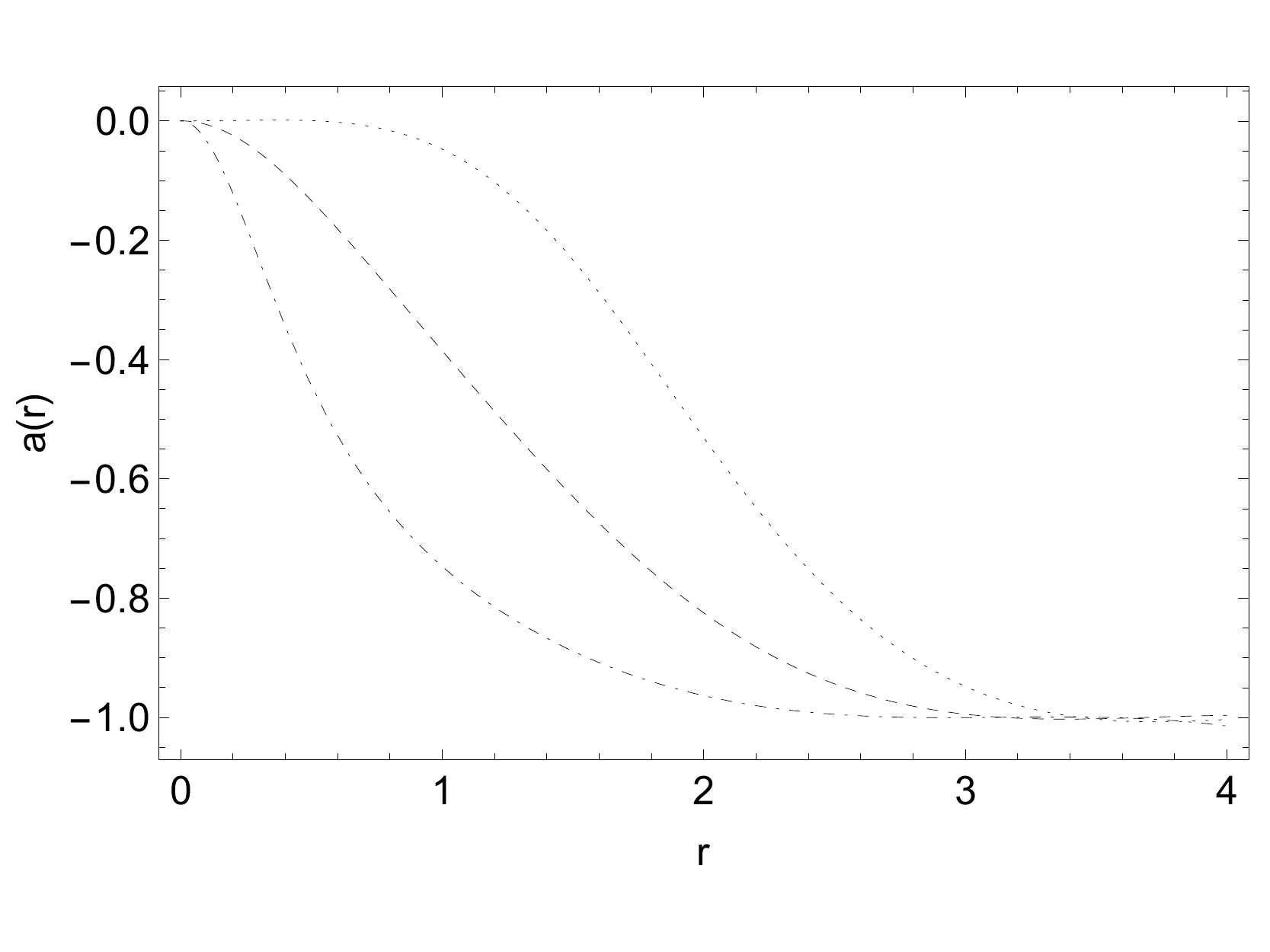}%\hspace{1cm}
    \includegraphics[height=5cm,width=7.5cm]{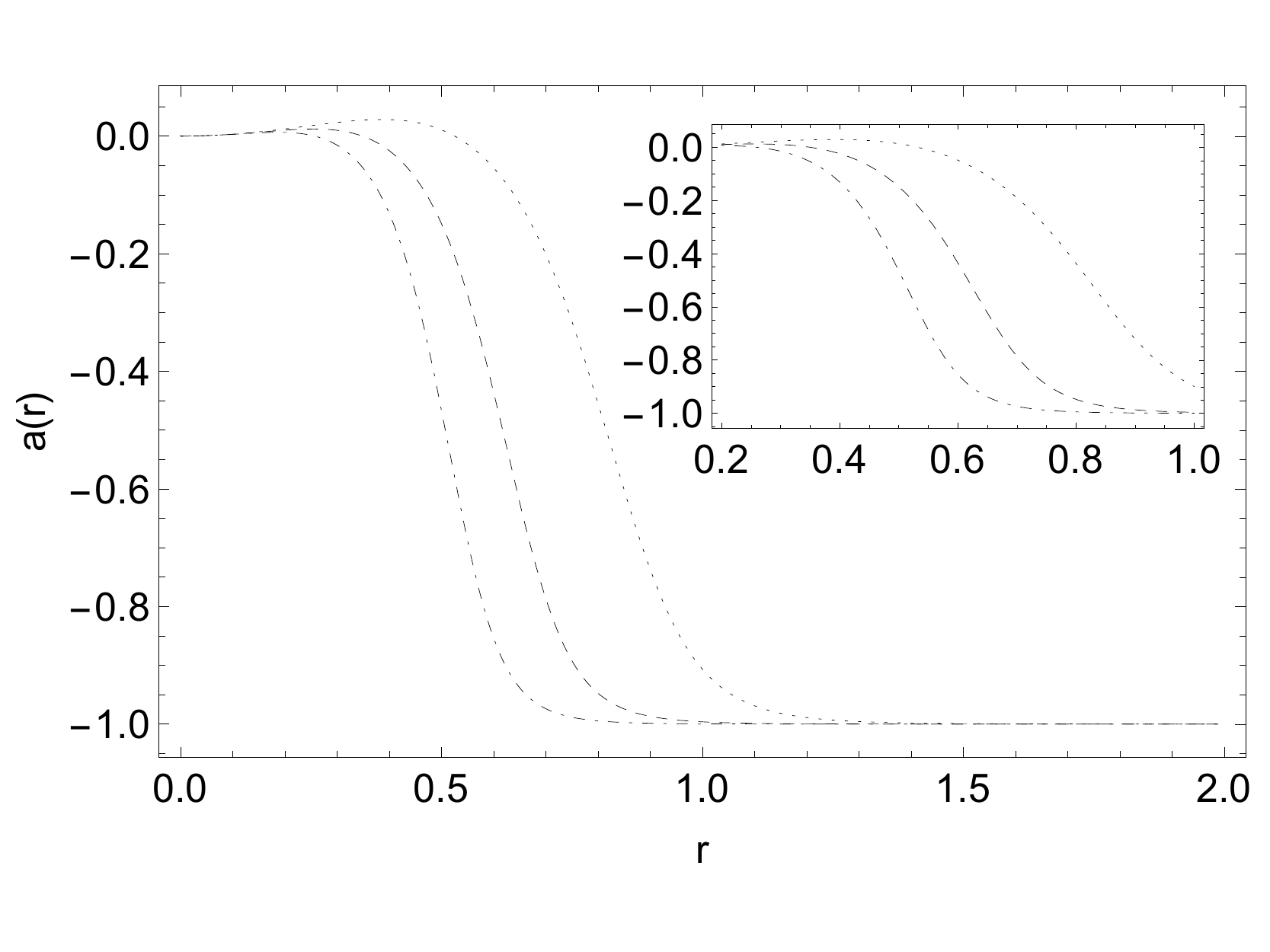}\\ \vspace{-0.5cm}
    \hspace{1cm}(a) \hspace{7.5cm} (b)
    \vspace{-0.5cm}
    \caption{(a) Numerical solutions of the variable field $a(r)$ with winding number $N=1$ and $m=1$. (b) Numerical solutions of the variable field $a(r)$ with winding number $N=1$ and $m=2$.}
    \label{fig5}
\end{figure}

The solutions obtained for the field variables in the pure model seem to behave as field configurations with compact-like profile, since the fast growth of the field is evident until reaching the vacuum expected value. Consequently, after reaching the vacuum expected value the model solutions remain stable. We remark that the compact-like profiles that we refer to arise in the study of classical field solutions are configurations that quickly reach the vacuum state (in a finite region) and remain stable after reaching the theoretical vacuum. Due to this characteristic, they are called compact-like structures. This type of structure has been intensively studied in other works, see f. e. Refs. \cite{LDA,CDC,FCCA1}. On the other hand, numerical solutions of the gauge field are shown in Fig. \ref{fig5}.

We see that the chosen generalization, $\mathcal{F}=\cos^m f(r)$, intensifies the fluctuations of the $\Phi$ field, since the theory potential remains unchanged. The variation of the parameter $m$ allows the deformation of the structure, taking it from a kink-like solution to a compact-like configuration.

\begin{figure}[ht!]
    \centering
    \includegraphics[height=5cm,width=7.5cm]{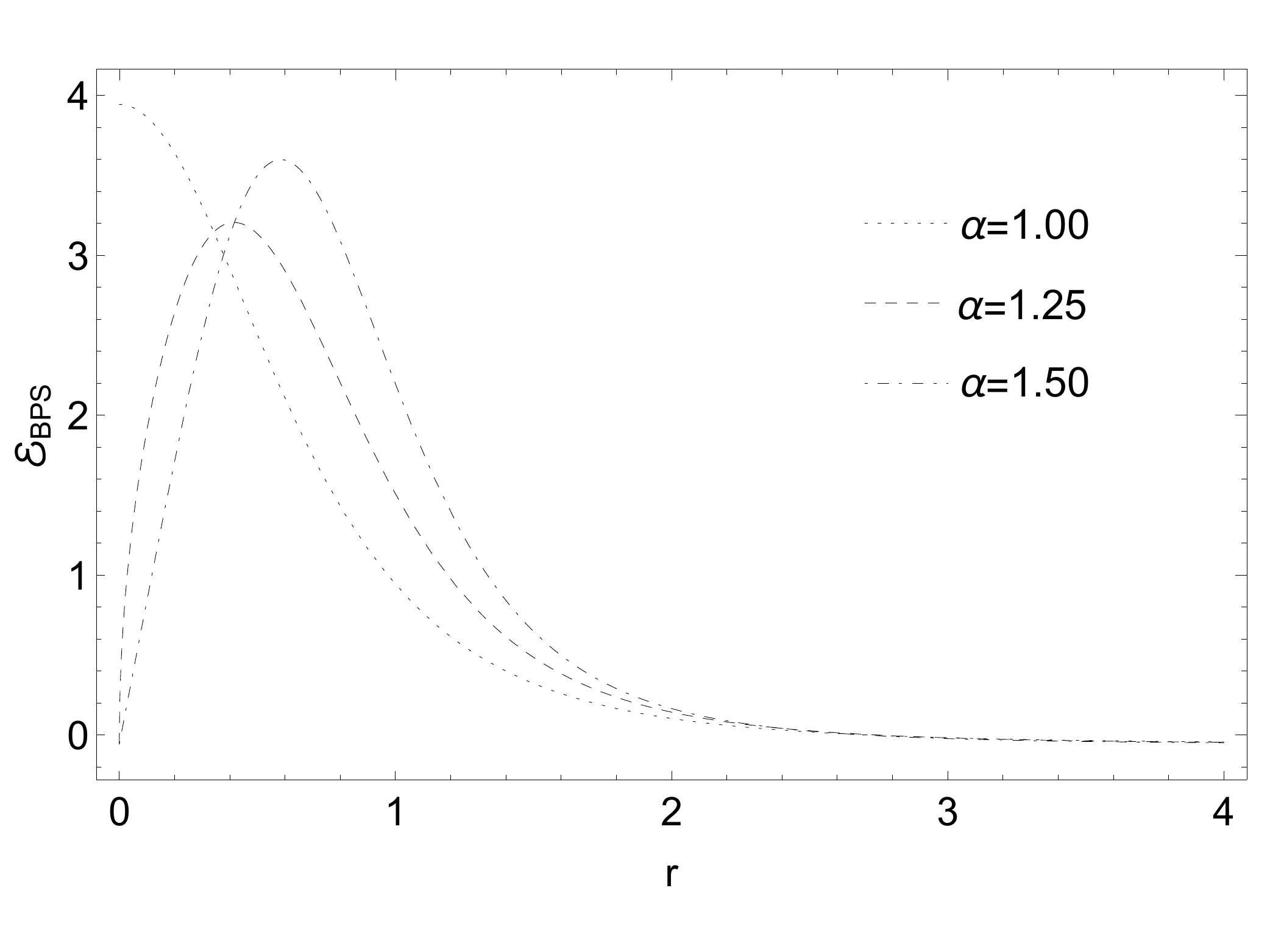}
    \vspace{-0.5cm}
    \caption{Energy density for the cases $m=1$ and $m=2$ with several values of the parameter $\alpha$.}
    \label{fig6}
\end{figure}

By analyzing the simulation (Fig. \ref{fig4}), we see that for $m>2$, the structures will have infinite energy settings, so they will not serve our purpose. In other words, for $m>2$, we do not obtain topological solitons. It is still interesting to note that the configurations of the field variables $f(r)$ will have a kink-like profile when $0\leq m\leq1$, and for $1<m\leq 2$ compact-like configurations. Using the numerical interpolation method with steps of $10^{-4}$, it is concluded from the simulation results that for $m$ outside the range $[0,2]$ the topological solitons are not detected, i. e., we were unable to find field configuration of finite energy and integer topological charge for the boundary (\ref{cont}). This is because the equation of the decoupled field variable $f(r)$ has the contribution of the term $\frac{m}{2}f^{_{'}{2}}(r)\sin^{-1 -\frac{m}{2}}f(r) \cos^{-\frac{m}{2}}f(r)$. This term for $m>2$ dominates over the other terms, so that the localized energy configurations do not exist for the boundary (\ref{cont}). Also, for $m\leq 2$ the contribution of the term is small allowing to obtain the solutions shown in Fig. (\ref{fig4}) with energy displayed in Fig. (\ref{fig6}).

As expected, despite the peculiar profile of the fields in this topological sector, the structures are described by finite energy configurations and have a shape similar to solitons of a KdV-like theory \cite{Pelinovsky}. This feature is clear in Fig. \ref{fig6}.

A peculiar characteristic of the structure of the pure model appears when we investigate the energy configurations for other values of the parameter $m$. By interpolation it is observed that regardless of parameter $m$ (Eq. (\ref{pure})), the energy profile remains invariant. It is interesting to mention that, in this case, the energy is also unchanged under a modification of the parameter $\alpha$. Due to these peculiarities, the numerical solution of the BPS energy density shown in Fig. \ref{fig6} is the same in both cases, i. e., for $m=1$ and $m=2$.

\begin{figure}[ht!]
    \centering
    \includegraphics[height=5cm,width=7.5cm]{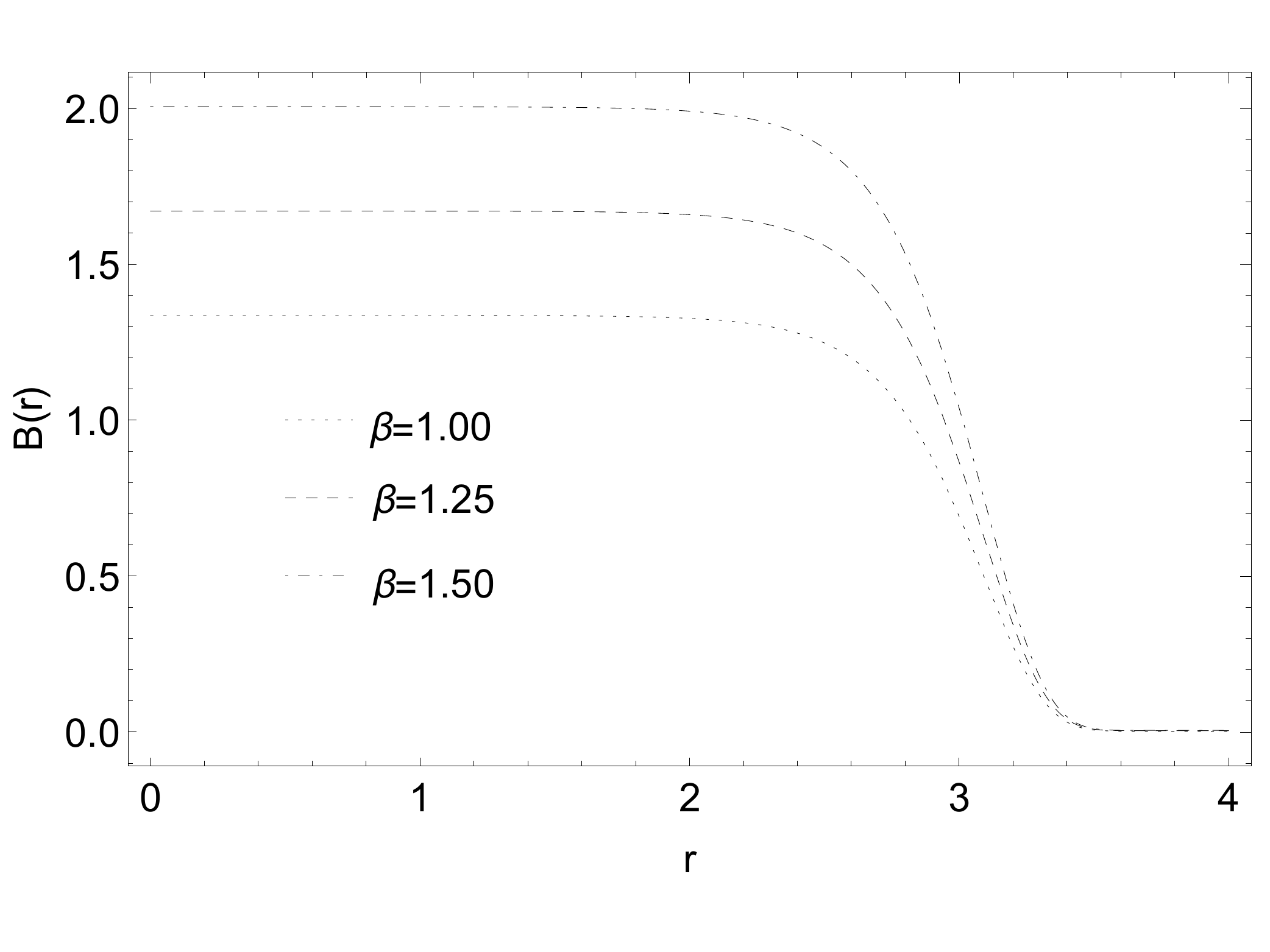}%\hspace{1cm}
    \includegraphics[height=5cm,width=7.5cm]{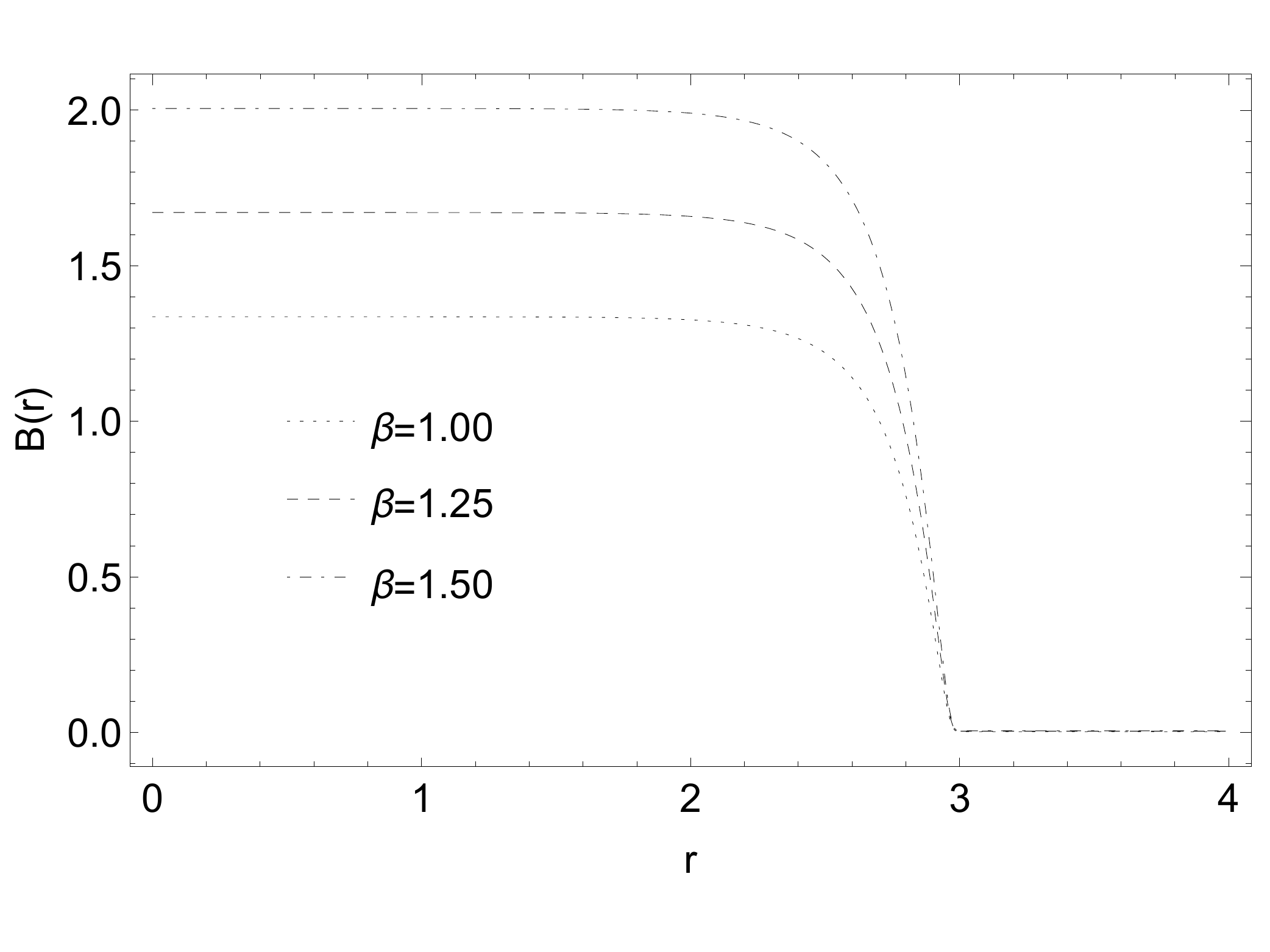}\\ \vspace{-0.5cm}
    \hspace{1cm}(a) \hspace{7.5cm} (b)
    \vspace{-0.5cm}
    \caption{(a) Magnetic field of the vortex when $N=1$ and $m=1$. (b) Magnetic field of the vortex when $N=1$ and $m=2$.}
    \label{fig7}
\end{figure}

To finalize the analysis of the topological vortices of this model, allow us to investigate the magnetic field produced by the non-minimal vortex. We found the magnetic field by numerical simulation, using Eq. (\ref{cmagnet}). Thereby is exposed in Fig. \ref{fig7} the numerical profile of the magnetic field.

In the pure model, the dielectric permeability of the vortex is $\mathcal{G}=U$. In this case, the vortex has energy larger around the vortex core. Nonetheless, the vortex profiles are modified to change the parameter $m$. This change shifts the maximum energy point of the vortex core (when this occurs, a vortex with ring-like shape appears). It seems to us that this peculiar behavior in this model is due to the shape of the magnetic field that these structures carry. The magnetic field is directly related to the new class of solutions of field variables, i. e., the solutions type KdV-solitons that arise when $m$ is large. To finish this brief discussion of the pure model, it is worth mentioning that in the limit of $m\to 0$ and $\mathcal{G}\to 1$, the results of the canonical model (Sec. 3.1) are retrieved.

\subsection{The mixed non-canonical model: $\mathcal{F}=\psi(r)$}

We exposed in the previous sections that the $\phi^4$ theory admits non-minimal vortex solutions in the non-generalized case (Sec. I) and in the pure case (Sec. II). Now, allow us to study the generalized model by a function $\psi$ (i.e., a mixed non-canonical model). The simplest case is when we assume $\mathcal{F}=\psi(r)$. In this condition, the BPS equations are
\begin{align}\label{bps17}
f'(r)=\pm\frac{N}{r\sqrt{\tanh(\text{ln}(r^\alpha))}}[a(r)-1]\sin f(r),
\end{align}
and
\begin{align}\label{bps18}
a'(r)=\pm\frac{r}{N}\cos f(r)\bigg[1-\frac{1}{3}\cos^2 f(r)\bigg].    
\end{align}
As we are working on a model with $O(3)$ symmetry, the field variable $f(r)$ and their derivative $f'(r)$ must have a positive-defined profile. It happens so that the topological structures have the properties of a kink-like solitary wave. However, when $r>0$ for some values of $\alpha$ the function $f'(r)$ will not be positive. In this case, if the structure is not positive-defined, it will not serve our purpose, i. e., it will not describe kink-like structures.

\begin{figure}[ht!]
    \centering
    \includegraphics[height=5cm,width=7.5cm]{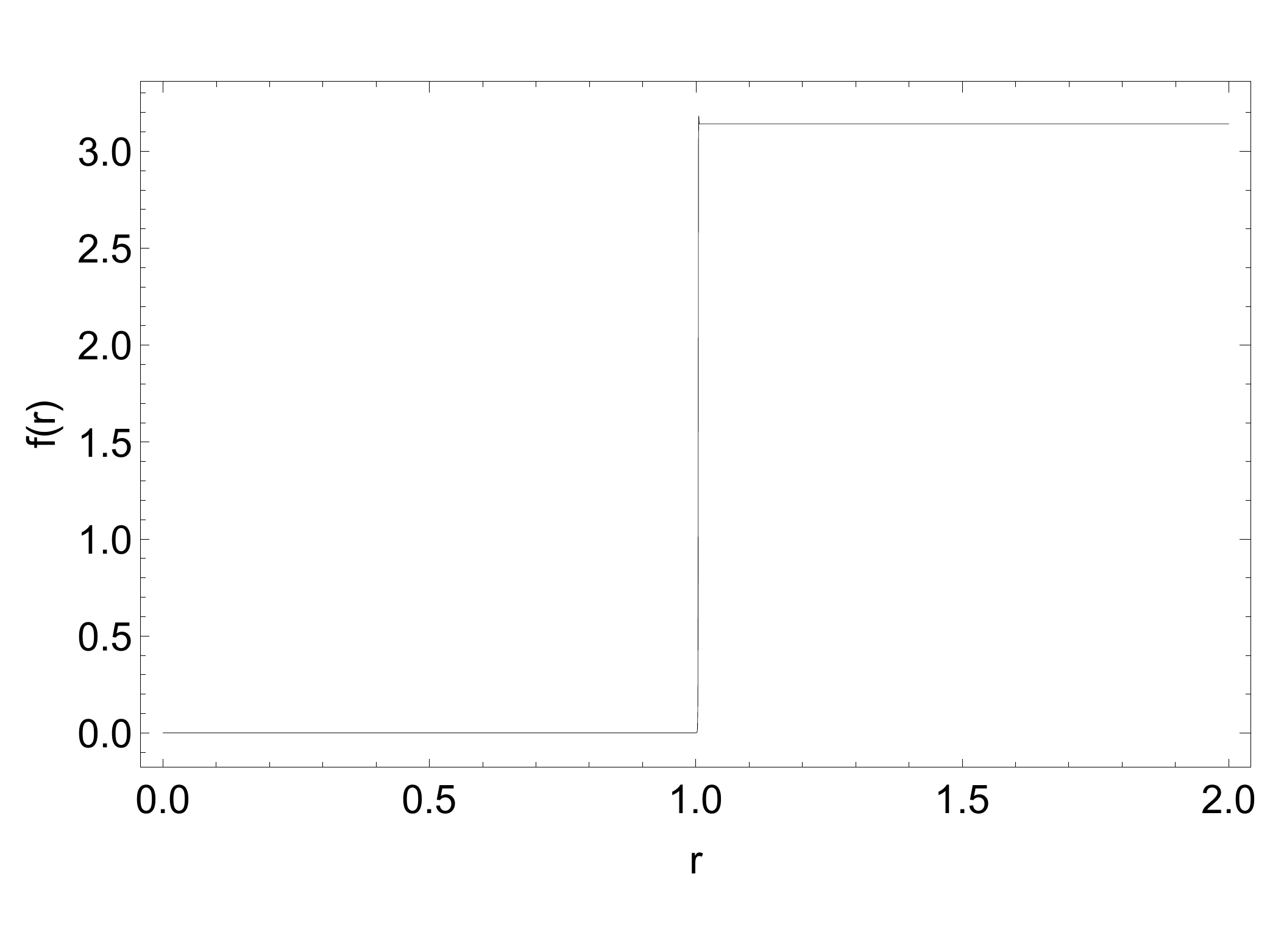}
    \vspace{-0.5cm}
    \caption{Configuration of the variable field $f(r)$.}
    \label{fig8}
\end{figure}

As in the previous cases, let us investigate the solutions of the variable field $f(r)$. In this preliminary investigation, we observed a peculiar behavior in the sector of the O(3)-sigma model with the non-canonical generalized term described by a coupling of the kinetic term of the sigma model with the field $\psi(r)$ given by Eq. (\ref{psi}). Indeed, it has limited field configurations, i. e., the variable field settings assume a profile similar to the step function. It is clear from Fig. \ref{fig8}, that this field profile is a limit behavior of a kink solution centered on $r=1$. This behavior seems to us to be a consequence of the mixed non-canonical case having a mixture of the $\Phi$ and $\psi$ fields. In other words, the kink of the topological sector $\psi $ interferes on the sector of the O(3)-sigma model so the only possible solution for this field configuration would be a field with the profile of the step function (note that the solution in the topological sector of the O(3)-sigma seems to be a case limited by sector $\psi$). It is still perceived that the kink of the $\psi$ sector interferes in order to locate the solitons of the sector of the O(3)-sigma model. In this way, the interference leads to only non-physical vortices step-like, that is, field configurations with infinite energy.

As an immediate consequence of this profile of the variable field $f(r)$, we notice that due to Eq. (\ref{bps17}) the magnetic field explodes in the near of the change of repeated field, i. e., $r=1$. Similarly, the topological structures of the mixed model will have an infinite energy density since the energy density of the model tends to diverge due to the contribution of term $f'(r)$.

Here it is interesting to mention that this problem can be solved by assigning other forms to the function $\psi$, since this result is a consequence of the location of the field $\psi$ around $r=1$ (point where the kink structure of the sector $\psi$ is located).

\section{Conclusions}

In this work, we studied the topological vortex solutions governed by the dynamics of a scalar field and the O(3)-sigma model coupled to the Maxwell field in a non-minimal way. For the generalized model in Sec. 3, the vortices with spherically symmetric configurations assume different profiles. Depending on the shape of the generalization function, the solutions can be from the kink solutions until its limit case, the variable field with a profile step-like function.

Analyzing each topological sector, we notice that the $\psi(r)$ sector solutions are the kink-like solutions. On the other hand, in the $\Phi$ sector, we observe three initial possibilities for the emergence of a new class of structures. The possibilities are the usual canonical model (in the case $\mathcal{F}=1$), the pure case, and the mixed case. The pure case happens when the non-canonical kinetic term is generalized by a contribution from the field $\Phi$. At the same time, the mixed model generalizes the topological sector of the sigma model with a function derived from the sector $\psi(r)$.
%Analyzing each topological sector, we note that the solutions of $\psi(r)$-sector are the kink-like solutions. Meanwhile, in the $\Phi$-sector, we observe three initial possibilities of the emergence of a new class of structures. Among them, there has been the usual canonical model (in case $\mathcal{F}=1$ ), the pure case with a non-canonical kinetic term generalizing the kinetic contribution of the $\Phi$ field as a function of one of the triplet components of the sigma model field. Finally, we have the mixed-model that generalizes the topological sector of the sigma model with a function derived from the $\psi(r)$-sector.

From an overview, the appearance of vortices with energy density related to the parameter $\alpha$ is evident. For higher values of parameter $\alpha$, the energy density of the model will tend to locate around the center of the kink. On the other hand, the parameter $\beta$ from the O(3)-sigma model is responsible for changing the intensity of the vortex's magnetic field, thus modifying the magnetic flux. It is interesting to mention that all vortices found have quantized magnetic flux due to the winding number.

In the canonical model, we can see that regardless of the parameter $m$ (Eq. (\ref{pure})), the energy profile remains invariant. It is interesting to mention, in this case, that the energy also remains unalterable under the alteration of the parameter $\alpha$. Due to these numerical solutions peculiarities, the BPS energy density of the vortices displayed in Fig. \ref{fig6} is the same for $m=1$, and $m=2$. We observe that by modifying the parameter $\alpha$, the vortex may have a higher energy profile. Therefore, the flux of energy radiated by the structure is more intense if $\alpha$ is large.

An attractive result appears in the mixed case, i. e., when we couple the generalization term of the $\psi(r)$ to the kinetic part of the sigma-O(3) model. In this case, the configurations of the variable field $f(r)$ assume a profile similar to the step function. This behavior seems to be a consequence of the mixed non-canonical case having a mixture of the $\Phi$ and $\psi$ fields from Eq. (\ref{psi}). Therefore, the kink of the topological $\psi$-sector is interfering (through the generalization term) over a sector of the O(3)-sigma model. So that the only possible solution for this field configuration, in this situation, will be a field with the profile of the step-like function. It is worth remarking that this profile of $f(r)$ is a non-physical topological class since the structures coming from this configuration have infinite energy.

A perspective of this study is to understand how these structures behave in curved spacetime. Not far away, another immediate possibility is the extension of the model to a theory with Lorentz symmetry breaking. We hope to carry out these studies in future work.
%------------------------------------------------------------------------
\section*{Acknowledgments}

The authors thank the Conselho Nacional de Desenvolvimento Cient\'{i}fico e Tecnol\'{o}gico (CNPq), Grant No. 309553/2021-0 (CASA), and Coordena\c{c}\~{a}o do Pessoal de N\'{i}vel Superior (CAPES) for financial support. We are grateful to the anonymous referee for his valuable criticism, suggestions, and comments.

\end{document}